\documentclass[letter]{aa}  
\usepackage[normalem]{ulem}

\usepackage{graphicx}
\usepackage{txfonts}

\begin{document}

   \title{Manifestation of the stellar wind cycle in \\
   infrared images. Example of the heliosphere}

   \author{ 
            I.P. Zabolotnyi\inst{\ref{inst1},\ref{inst2}}
            \and
            V.V. Izmodenov\inst{\ref{inst1},\ref{inst2},\ref{inst3}}
          }

   \institute{
                Space Research Institute of Russian Academy of Sciences, Profsoyuznaya 84/32, Moscow 117997, Russia\\
                \email{ivan.zabolotniy@cosmos.ru} \label{inst1}
                \and
                Lomonosov Moscow State University, Leninskie Gory 1, Moscow 119991, Russia\label{inst2}
                \and
                HSE University, 20 Myasnitskaya Ulitsa, Moscow 101000, Russia\label{inst3}
             }

   \date{}

  \abstract
   {}
   {The regions in which stellar winds interact with the interstellar medium, also known as astrospheres, can be observed in detail through the thermal emission of the interstellar dust particles, resided in plasma. Interstellar dust is also directly observed in the vicinity of the Sun with dust detectors onboard spacecraft, and it is known to be affected by the interplanetary magnetic field. The main goal of this work is to show how the change in the interplanetary magnetic field with the solar cycle affects the infrared picture of the heliosphere.}
   {To compose a synthetic intensity map of the interstellar dust thermal emission, we used the Monte Carlo method to calculate the distribution of the dust particles inside the heliosphere. We considered the effects of the heliosphere boundaries and the non-stationary current sheet.}
   {The change in the Parker magnetic field caused by the solar activity cycle leads distinguishable features in the mid-infrared emission maps of the heliosphere. The distribution of the interstellar dust in the vicinity of the Sun we calculated suggests that small particles linger outside of the heliosphere, and medium-size particles are mostly affected by the changing interplanetary magnetic field, which leads to number density waves in the tail region of the heliosphere. Finally, large particles form a bulge behind the Sun.}
   {}


   \maketitle

\section{Introduction}
    
    The region of the solar wind is filled with plasma that expands from the surface of the Sun into the interstellar medium (ISM). Since the solar wind becomes supersonic at small distances and the Sun travels in the ISM at a speed that exceeds the local speed of sound, a complex structure arises that consists of two shock waves, the heliospheric shock wave (TS) and the bow shock (BS), and of  tangential discontinuity, the heliopause (HP) \citep{Baranov1970}. This structure is called heliospheric boundary layer or shock layer. The plasma convects the interstellar magnetic field in the region outside of the HP. Inside the HP, it convects the solar magnetic field, forming the Parker spiral \citep{Parker1958}.

    Interstellar dust particles enter the heliospheric boundary layer and move under the effects of the gravitational attraction to the star, radiative pressure from stellar photons, the electromagnetic force, and the drag force. Smaller particles linger outside of the HP, while larger particles penetrate into the heliosphere. In- situ observations on spacecrafts, such as Ulysses \citep{Grun1993} and Stardust \citep{Westphal2014}, confirmed the presence of interstellar dust particles in the vicinity of the Sun. The dynamics of the dust particles that move inside the heliosphere are affected by the interplanetary magnetic field (IMF)\citep{Sterken2022,Baalmann2025}, which itself has a dynamic structure.
    
    The solar wind changes periodically with the 11-year activity cycle of the sunspots. During this cycle, the solar magnetic field changes polarity, which leads to the 22-year magnetic cycle and to alteration of the IMF. Since interstellar dust dynamics inside the heliosphere are governed by the IMF, the distribution of the interstellar dust particles in the heliosphere is affected by the solar activity \citep{Godenko2024a}.
    
    Some other stars similar to the Sun are known to have a magnetic activity cycle. The chromospheric magnetic activity of distant stars is observed through the variation in the S index, which is an emission measure of the Ca II H and K lines \citep{Baliunas1995}. The challenging part of these observations is that they need to be conducted on long timescales. The Mount Wilson observations \citep{Wilson1978} collected one of the longest datasets on stellar activity over several decades. A number of datasets of long-term stellar activity observations are available today (see \cite{Jeffers2023} for a recent review). Nevertheless, only a few stars are confirmed to experience the polarity change at activity maxima like the Sun (\cite{Bellotti2025}). The first stars to show a polarity change were Cygni A \citep{Boro2016} and $\tau$ Bootis \citep{Jeffers2018}, with 7.3 yr and 10 yr cycles, respectively.
    
    The shock layers of other stars are observed in the mid-infrared through the thermal radiation of interstellar dust particles that penetrate this shock layer and are heated by the star \citep{Cox2012, Kobulnicky2016,Gvaramadze2020,Larkin2025,Russeil2025}. Some stars release dust into the ISM, and this dust can form circular shells around the stars \citep{Drudis2020}. The interplanetary dust environment of the Sun is relatively dust poor in comparison to debris discs of other stars \citep{Poppe2019}, which, however, can make the interstellar dust prominent on the scale of the heliosphere.
    
    Most current models of the interstellar dust distribution inside the heliosphere are intended to analyse the Ulysses data \citep{Godenko2024b, Baalmann2025}. \cite{Frisch2013} modelled the large-scale distribution of dust particles in the heliosphere using the heliosphere model of \cite{Pogorelov2008}. They noted that large dust grains that penetrate the heliosphere are concentrated either at the solar equatorial plane during the focusing phase or in the polar region during the defocusing phase, but the authors only considered a stationary IMF. \cite{Alexashov2016} modelled the interstellar dust distribution with a non-stationary IMF, but mainly focused on the deflection of the dust at the heliosphere boundaries.
    
    The goal of this Letter is to answer the question how the stellar cycle manifests itself in the infrared image of the astrosphere. For this purpose, we studied the interstellar dust distribution in the vicinity of the Sun and its effect on the infrared picture of the heliosphere. We used a numerical model of the interstellar dust distribution with a Monte Carlo approach that included the effect of the heliosphere boundaries and non-stationary IMF to compose synthetic mid-infrared emission maps of the heliosphere. The structure of the paper is as follows. In Section 2 we briefly describe the model. In Section 3 we present and discuss our results. Section 4 concludes this paper.

\section{Methods}

    Interstellar dust particles are heated by the radiation of the Sun and emit thermal radiation. The emissivity of the potpourri of dust particles of different sizes can be expressed in the following form \citep{Katushkina2019}:
    \begin{equation}
        j_{\nu} = \int \limits_{a_{min}}^{a_{max}}
        Q_{abs}\left( \nu \right) \pi a^{2}
        B_{\nu} \left( T_d \right) \frac{dn_d}{da} da,
    \end{equation}

    where $a$ is the dust particle radius, $Q_{abs}$ is an absorption coefficient of the dust particle, $B_{\nu}$ is the Plank function, $T_d$ is the dust particle temperature, $dn_d/da$ is the number density of the particles of sizes from $a$ to $a+da$, and $\nu$ is the wave frequency.
    
    The spectral intensity from the dust grains can be derived by integrating the spectral emissivity along the line of sight,
    \begin{equation}
        I_{\nu} = \int \limits_{s_{min}}^{s_{max}} j_{\nu} ds,
    \end{equation}

    where s is the coordinate along the line of sight.
    
    Typically, the thermal radiation of dust particles in the vicinity of other stars is observed in the mid-infrared. We chose $\nu_0$, which corresponds to the spectral emission maximum at the wavelength $\lambda_0 = 50\mu m$. We show intensity maps that represent the integrated emissivity along lines of sight parallel to the direction towards the observer and evenly spaced in the projection plane.
    
    The absorption coefficient of the dust particles was calculated using Mie theory \citep{Bohren1998}. Since there is no in situ evidence of PAH and carbon particles inside the heliosphere, we restricted our calculations to the astronomical silicates. We employed the values of the absorption coefficient for the astronomical silicates obtained by \cite{Draine2003}.
    
    The temperature of the dust particles was obtained via energy  equilibrium between absorption of the interstellar and solar radiation field photons and emission of a single particle,
    \begin{align}\label{temperature}
        \int \limits_{0}^{\infty} 4 \pi^2 a^2 Q_{abs}
        B_{\nu} \left( T_d \right) d \nu = &
        \nonumber \\ 
        = \int \limits_{0}^{\infty} \pi^2 a^2 Q_{abs}
        \left( \frac{R_{\odot}}{r} \right)^2 B_{\nu} \left( T_{\odot} \right) d \nu &+ 
        \int \limits_{0}^{\infty} 4 \pi^2 a^2 Q_{abs} ,
        J_{\nu} d \nu
    \end{align}

    where $R_{\odot}$ is the radius of the Sun, $T_{\odot}$ is the temperature of the Sun, $r$ is the distance to the Sun, and $J_\nu$ is the intensity of the isotropic interstellar radiation field.

    The size distribution function of the dust particles was assumed to be homogeneous in the ISM,
    \begin{equation}
        \frac{dn_{d,ISM}}{da} = C_{ISM} F_{ISM}(a),
    \end{equation}
    
    where $F_{ISM}$ is the dust size distribution, and $C_{ISM}$ is the factor that keeps the interstellar dust to the plasma mass density ratio $\rho_d/\rho_p = 1/100$.

    To obtain the number density of the particles in the region that is disturbed by the Sun (i.e. inside the heliosphere boundary layer), we calculated the number density of particles of  distinct size $\hat{n}_d$, with the number density in the undisturbed medium set to unity. Then the number density inside the boundary layer was obtained as
    \begin{equation}
        \frac{dn_{d}}{da} = \hat{n}_d (a,\textbf{r}) C_{ISM} F_{ISM}(a),
    \end{equation}
    
    where $\textbf{r}$ is the position of the particle, and $\hat{n}_d$ is the number density of the particles of distinct sizes.

    \begin{figure}
        \centering
        \includegraphics[width=6cm]{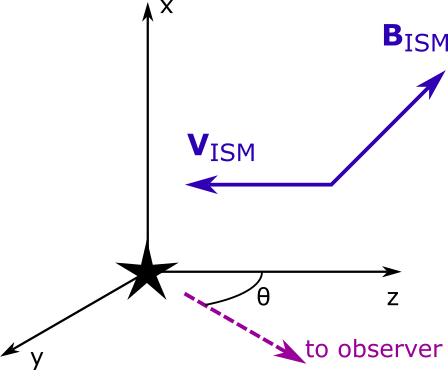}
        \caption{Coordinate system used for calculations. The Sun is at the origin, $V_{ISM}$ is the relative velocity of the interstellar dust and plasma, $B_{ISM}$ is the interstellar magnetic field, and $\theta$ is an angle between the direction towards the observer and the velocity of the ISM.}
        \label{Figure1}
    \end{figure}
    
    To calculate the number density of particles of distinct sizes, we used the Monte Carlo method described in detail by \cite{Godenko2024b}. We introduced the coordinate system so that the Z-axis was opposite to the direction of the velocity of the ISM, the X-axis was directed so that the homogeneous interstellar magnetic field vector lay in the XZ plane (BV plane), and the Y-axis completed the right-handed coordinate system (Fig. \ref{Figure1}).
    
    The Monte Carlo calculations were carried out in a rectangular box 1000 a.u. x 1000 a.u. x 1000 a.u., centred at the Sun. We used a regular rectangular grid, with 10 a.u. x 10 a.u. x 10 a.u. resolution in space and a one-year resolution in time. The boundary condition was set far from the heliosphere boundary layer in the undisturbed ISM, where all particles are assumed to move with the bulk velocity of the plasma. The plasma parameters and magnetic field strength inside the boundary layer were taken from a global MHD-kinetic model of the heliosphere \citep{Izmodenov2015, Izmodenov2020}. We considered a non-stationary heliospheric current sheet that propagates with plasma inside the heliosphere \citep[see][Appendix A]{Godenko2024b}.
    
    For simplicity, the dust particles were assumed to have a constant mass and spherical shape. The charge of the particles was calculated according to the balance between the sticking of thermal plasma protons and electrons, secondary electron emission, photoelectric emission, and the effect of cosmic rays \citep[see][]{Godenko2023b}.

    The number density of the interstellar dust inside the heliosphere was calculated for astronomical silicates of sizes in between $a_{min} = 10 \ nm$ and $a_{max} = 1 \ \mu m$. We used a standard power-law size distribution \citep{Mathis1977} to model the thermal emission maps.

\section{Results and discussion}
    
    \begin{figure}
        \centering
        \includegraphics[width=8cm]{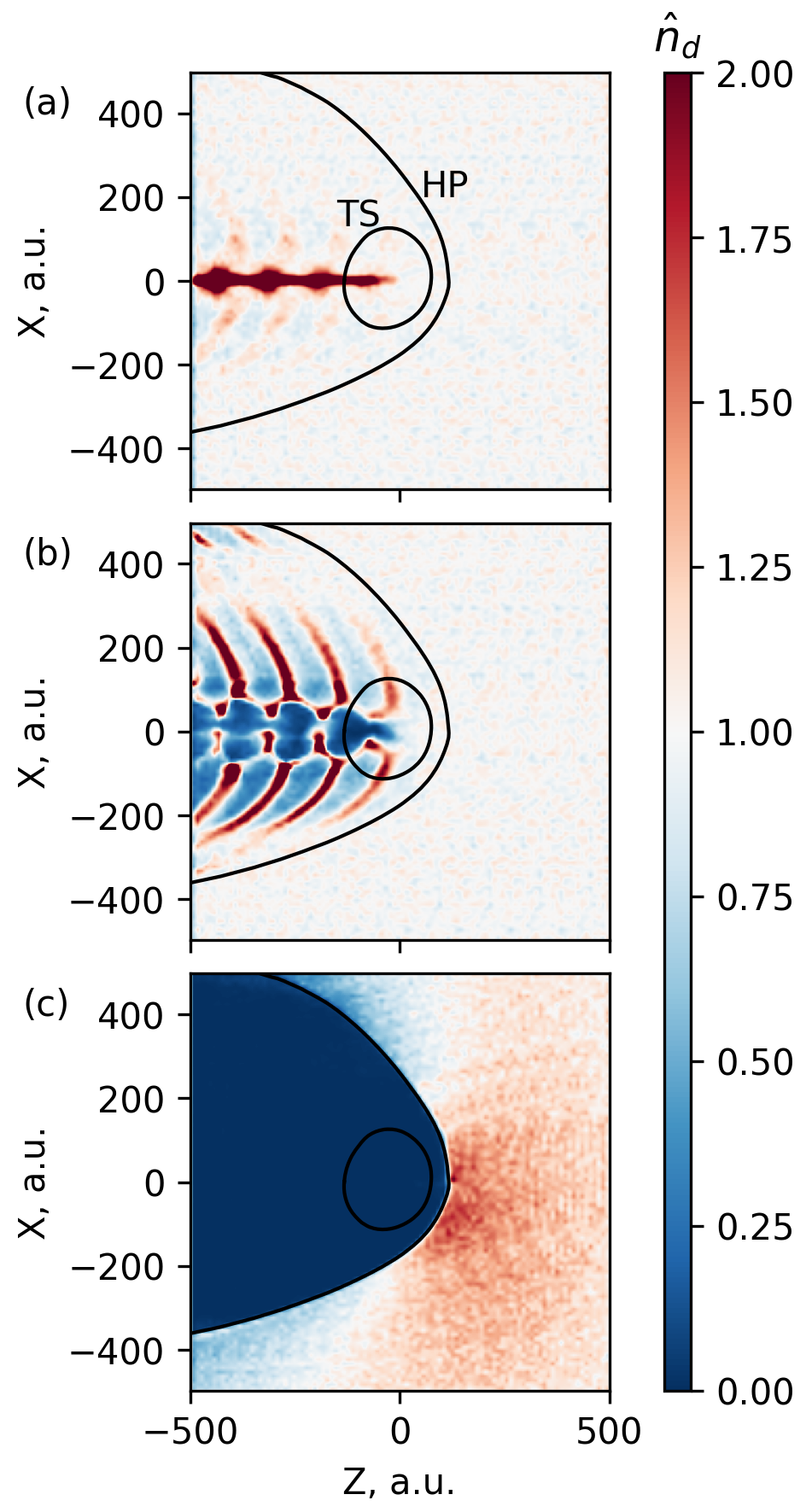}
        \caption{Resulting number density for the interstellar dust particles with sizes of 1000 nm (a), 300 nm (b), and 10 nm (c) in the BV plane. The Sun is at the origin. The solid white lines depict the position of the HP (outer parabolic line) and the TS (inner circular line).}
        \label{Figure2}
    \end{figure}
    
    Fig. \ref{Figure2} shows the number density of particles of different sizes in the BV plane. Small particles (i.e. particles smaller than $\approx 100$ nm) do not penetrate the heliosphere and stay in the outer shock layer (Fig. \ref{Figure2}a). This is consistent with in situ measurements onboard the Ulysses spacecraft.
    
    Medium-sized particles (e.g. 300 nm) penetrate the HP and are strongly affected by the IMF. As noted by \cite{Frisch2013}, for the stationary magnetic field, the number density of the particles that penetrate the heliosphere is higher in the polar regions of the Sun due to the Lorentz force. The non-stationary IMF causes number density waves in the tail region of the heliosphere (Fig. \ref{Figure2}b) because the magnetic field lines and the velocity of the plasma are tangential to the HP, resulting in the electric field normal to the HP. This electric field either pushes particles out or drags them inside the heliosphere, depending on the phase of the solar activity cycle. A cavity forms behind the Sun due to the combined effect of the radiation pressure (also known as $\beta$-cone) and the defocusing of the particles near the equatorial plane of the Sun \citep[see e.g.][]{Godenko2023a}.
    
    Since the electromagnetic force is related to the size of the particle approximately as $a^{-2}$, large particles are mainly affected by gravity and follow the hyperbolic trajectories. This causes the number density bulge behind the Sun \citep{Landgraf2000}.
    
    \begin{figure*}
        \centering
        \includegraphics[width=18cm]{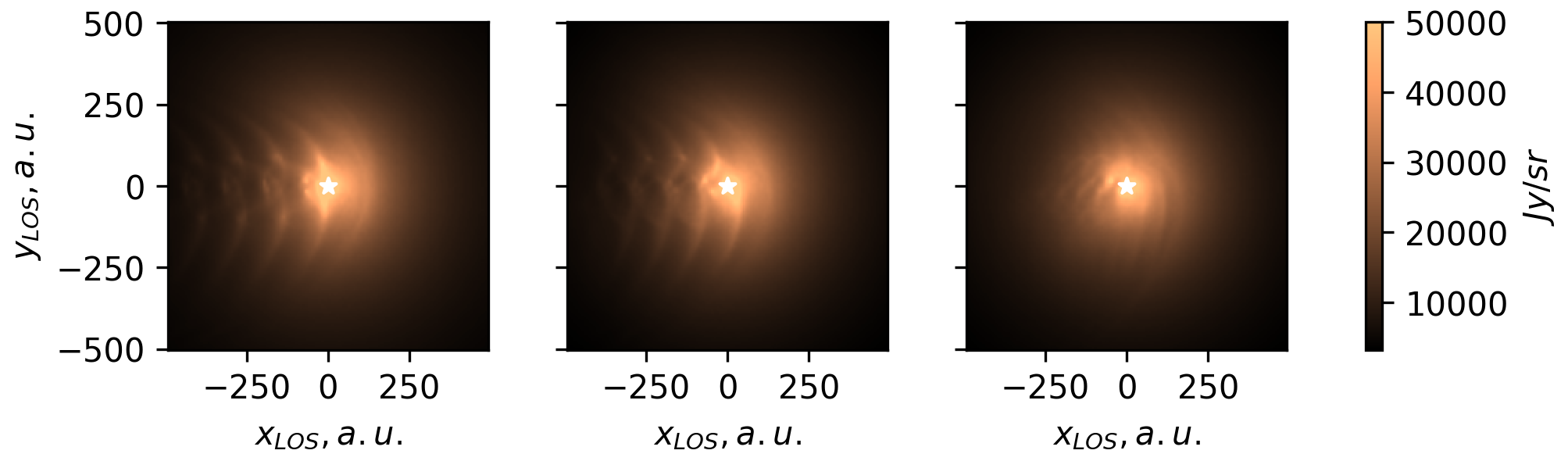}
        \caption{Synthetic intensity maps of the thermal emission of interstellar dust (at   wavelength of $50 \mu m$) in the vicinity of the Sun. The angle between the line of sight and the relative velocity vector of the Sun is $90^\circ$ (left), $60^\circ$ (middle), and $30^\circ$ (right). $x_{LOS}$ and $y_{LOS}$ stand for the local coordinates in the projection plane.}
        \label{Figure4}
    \end{figure*}
    
    \begin{figure}
    \centering
    \includegraphics[width=8cm]{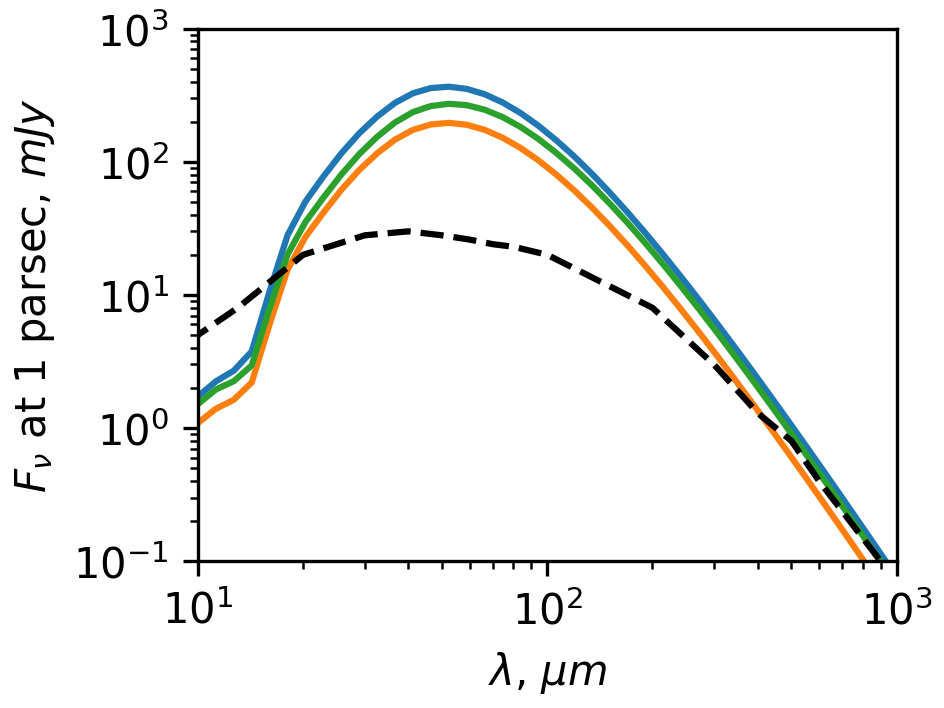}
    \caption{Spectral emission distribution of interstellar dust particles inside the heliosphere for three different size distributions of particles. Classical MRN (green) \citep{Mathis1977}, a truncated power law (blue) \citep{Mathis1996}, and the most recent (orange) \citep{Hensley2023}. The dashed line indicates the interplanetary dust SED from \cite{Poppe2019}.}
         \label{Figure3}
    \end{figure}
    
    The resulting synthetic mid-infrared images of the heliosphere from different angles between the direction towards the observer and the relative velocity of the Sun and ISM are shown in Fig. \ref{Figure4}. Particles that penetrate the heliosphere have a higher emissivity because their temperature is higher, which causes the wavy features in the synthetic intensity maps. The distance between the local intensity maxima corresponds to the solar cycle and the apparent relative velocity of the Sun as
    \begin{equation}
        l = t_{cycle} v_{ISM}\sin{\theta},
    \end{equation}

    where l is the distance between the intensity maxima, $t_{cycle}$ is the solar cycle period, $v_{ISM}$ is the velocity of the Sun relative to the local ISM, and $\theta$ is an angle between the line of sight and the direction of the relative velocity.

    When the relative velocity vector lies in the projection plane, the distance between local emission maxima is $l \approx 120\ a.u.$. With decreasing $\theta$, the waves are smeared together and become almost indistinguishable at $\theta \approx 30^\circ$. The faint arc in front of the star is the emission of the ISD particles outside of the HP.

    We also calculated intensity maps and the spectral emission distribution (SED) for an observer at 1 parsec  using the MRN size distribution,the  truncated power-law size distribution \citep{Mathis1996}, and the size distribution and material composition by \cite{Hensley2023}. The changes in the intensity maps are almost indistinguishable, and the effect on the SED is rather quantitative, not qualitative (Fig. \ref{Figure3}) because the infrared image is mostly affected by medium-sized particles, for which the size distribution is relatively well constrained. Moreover, PAH or carbonaceous particles are relatively small and therefore stay in the outer shock layer.

    An actual image of the Sun would also include the emission of the debris disc, which is located near the ecliptic plane. \cite{Poppe2019} calculated the emission and SED of the modelled debris disc of the Sun. The SED of the interstellar dust exceeds that of the interplanetary dust at wavelengths  $\approx 20..300 \mu m$, with the flux maximum at $\approx 200 mJy$ near $50 \mu m$. The relative luminosity of the interstellar dust is $L_{ISD}/L_{\odot} \approx 7 \cdot10^{-7}$, and in the same order as the luminosity of the interplanetary dust, which suggests a strong effect of the interstellar dust on the infrared excess of the Sun.

    \section{Conclusions}
    We performed a Monte Carlo computation of the interstellar dust distribution in the global heliosphere using a numerical kinetic-MHD model of the heliosphere to investigate the effect of the changing IMF on the mid-infrared picture of the Sun. We list the results of our computations below.
    \begin{itemize}
        \item The medium-sized particle distribution inside the heliosphere is strongly affected by the IMF, which shapes number density waves through the change in the IMF during the solar magnetic activity cycle.
        \item These waves are visible in the thermal emission maps in the mid-infrared from different angles, up to $\approx30^\circ$ between the relative velocity and the direction to the observer. The distance between local emission maxima is related to the activity cycle period and apparent motion of the Sun relative to the local ISM.
    \end{itemize}

    Using the Sun as an example, we demonstrated overall that infrared images of the astrospheres might be sources of new unique information on the stellar activities.

    \begin{acknowledgements}
        Authors acknowledge support from the Russian Science Foundation grant 25-12-00240. Authors also thank Egor Godenko for providing the dust distribution in the heliosphere.
    \end{acknowledgements}
    
    \bibliographystyle{aa}
    \bibliography{biblio}

\end{document}